\documentclass[letterpaper]{article} % DO NOT CHANGE THIS
\usepackage{aaai25}  % DO NOT CHANGE THIS
\usepackage{times}  % DO NOT CHANGE THIS
\usepackage{helvet}  % DO NOT CHANGE THIS
\usepackage{courier}  % DO NOT CHANGE THIS
\usepackage[hyphens]{url}  % DO NOT CHANGE THIS
\usepackage{graphicx} % DO NOT CHANGE THIS
\usepackage{natbib}  % DO NOT CHANGE THIS AND DO NOT ADD ANY OPTIONS TO IT
\usepackage{caption} % DO NOT CHANGE THIS AND DO NOT ADD ANY OPTIONS TO IT
\usepackage{algorithm}
\usepackage{algorithmic}

\usepackage{physics}
\usepackage{subfig}
\usepackage{booktabs} 
\usepackage{xcolor}
\newcommand{\answerYes}[1]{\textcolor{blue}{#1}} 
\newcommand{\answerNo}[1]{\textcolor{teal}{#1}} 
\newcommand{\answerNA}[1]{\textcolor{gray}{#1}}

\usepackage{newfloat}
\usepackage{listings}
\DeclareCaptionStyle{ruled}{labelfont=normalfont,labelsep=colon,strut=off} % DO NOT CHANGE THIS
\floatstyle{ruled}
\newfloat{listing}{tb}{lst}{}
\floatname{listing}{Listing}
\title{Exploring Unknown Social Networks for Discovering Hidden Nodes}
\author{
    Sho Tsugawa, \textsuperscript{\rm 1}
    Hiroyuki Ohsaki\textsuperscript{\rm 2}
    \\
}
\affiliations{
    \textsuperscript{\rm 1}Institute of Systems and Information Engineering, University of Tsukuba\\
Tsukuba, Ibaraki 305--8573, Japan\\
    \textsuperscript{\rm 2}School of Science and Technology, Kwansei Gakuin University\\
1 Gakuenuegahara, Sanda, Hyogo 669-1330, Japan\\

    s-tugawa@cs.tsukuba.ac.jp, ohsaki@kwansei.ac.jp
}

\usepackage{bibentry}
\begin{document}

\maketitle

\begin{abstract}

In this paper, we address the challenge of discovering hidden nodes in unknown social networks, formulating three types of hidden-node discovery problems, namely, Sybil-node discovery, peripheral-node discovery, and influencer discovery. We tackle these problems by employing a graph exploration framework grounded in machine learning. Leveraging the structure of the subgraph gradually obtained from graph exploration, we construct prediction models to identify target hidden nodes in unknown social graphs. Through empirical investigations of real social graphs, we investigate the efficiency of graph exploration strategies in uncovering hidden nodes. Our results show that our graph exploration strategies discover hidden nodes with an efficiency comparable to that when the graph structure is known. Specifically, the query cost of discovering 10\% of the hidden nodes is at most only 1.2 times that when the topology is known, and the query-cost multiplier for discovering 90\% of the hidden nodes is at most only 1.4. Furthermore, our results suggest that using node embeddings, which are low-dimensional vector representations of nodes, for hidden-node discovery is a {\em double-edged sword}: it is effective in certain scenarios but sometimes degrades the efficiency of node discovery. Guided by this observation, we examine the effectiveness of using a bandit algorithm to combine the prediction models that use node embeddings with those that do not, and our analysis shows that the bandit-based graph exploration strategy achieves efficient node discovery across a wide array of settings.

\end{abstract}

\section{Introduction}
\label{sec:intro} 

Identifying individuals with specific characteristics in social networks is an important research issue across various domains.
For instance, methods have been studied for identifying social
media users who are interested in viral marketing campaigns~\cite{wang2013active,qiu2018deepinf}, influential users~\cite{tsugawa22:icwsm,zhao2020machine,panagopoulos2020influence}, individuals with specific political ideologies~\cite{wang2020heterogeneous,darwish2020unsupervised}, and malicious social bots~\cite{mendoza2020bots,fazil2021deepsbd,feng2021botrgcn,ferrara2016rise}.

Typically, the problem of identifying individuals with specific characteristics can be formulated as that of estimating node labels in a given social graph based on a supervised or semi-supervised machine learning (ML) framework.
Such problems have been studied independently for different types of target nodes to be identified, most studies assumed that the complete structure of the target social graph is available but that node labels representing attributes are only partially available~\cite{feng2021botrgcn,tsugawa22:icwsm,zhao2020machine,wang2020heterogeneous,mendoza2020bots}.
The goal is then to estimate the unknown node labels by using the partially available ones along with the social-graph structure. To estimate the node labels,
techniques such as label propagation~\cite{zhu2003semi} and graph neural networks~\cite{kipf2017semi} have been used, assuming known graph structure and labels for some nodes. Tasks such as Sybil detection~\cite{wang2017sybilscar,sun2020trustgcn}, influencer identification~\cite{tsugawa22:icwsm,zhao2020machine,panagopoulos2020influence,kim2023influencerrank}, and predicting nodes influenced by information diffusion~\cite{sankar2020inf} have also been studied under similar assumptions.

However, in the context of real-world scenarios involving the
identification of {\em hidden} or {\em hard-to-reach} populations within
a large-scale social network, obtaining the
topology of the target social graph is challenging~\cite{mccreesh2012evaluation,yadav2018bridging,Wilder18:influence}.
For instance, in domains such as epidemiology and social welfare, investigations have been undertaken into HIV-infected individuals~\cite{pando2012hiv,gile2011improved}, populations at high risk of sexually transmitted infections~\cite{pando2012hiv}, injecting drug users~\cite{heckathorn2002extensions}, and populations afflicted by poverty~\cite{heckathorn1997respondent}. In such studies, researchers use respondent-driven sampling (RDS)~\cite{heckathorn1997respondent,mccreesh2012evaluation} to explore the social graph to which these hidden populations belong, aiming to identify the target nodes. Through interviews and questionnaire surveys, researchers gather attribute information about the subjects (e.g., whether they are HIV-infected) and information about their friends. Subsequently, by repeating similar inquiries with these friends, researchers gradually obtain the structure of the social graph, with the aim of identifying the hidden populations of interest. Given the monetary and temporal costs associated with interviews and questionnaire surveys, it is desirable to uncover the target hidden populations with as few inquiries as possible.
Conversely, even in the realm of online social networks, where one may
envision scenarios involving the discovery of users with specific hidden
attributes (such as social bots~\cite{ferrara2016rise} or influencers~\cite{yadav2018bridging,Wilder18:influence,Mihara15:Influence}), the structure of the
social graph remains unavailable beforehand. For entities other
than social media operators, social graphs representing
relationships among social media users are typically accessible only
through application programming interfaces (APIs). Consequently, researchers resort to
using social media APIs to acquire user information while traversing
toward the target nodes. 
Given the monetary and temporal costs associated with API access, it
is desirable to uncover the target hidden users with as few API queries as possible.
These real-world scenarios motivate us to address the problem of discovering nodes with target labels in a social graph while exploring graphs with unknown topology~\cite{murai2018selective, wang2020heterogeneous, morales2021selective}.

In this paper, we address three types of tasks for discovering hidden target nodes in unknown social graphs.
First, we tackle the discovery of Sybil nodes~\cite{wang2017sybilscar,furutani2023interpreting,jia2017random,sun2020trustgcn}, which are malicious nodes within social networks, intent on spreading fake news or manipulating public opinion.
Second, we address the discovery of peripheral nodes in social networks; these are nodes with limited social capital, making them challenging to find through social ties.
Third, we address the discovery of influencers who can spread information to many other nodes in social networks~\cite{tsugawa22:icwsm,panagopoulos2020influence,kim2023influencerrank}.

\begin{figure}[t]
  \centering 
\includegraphics[clip,width=.8\columnwidth]{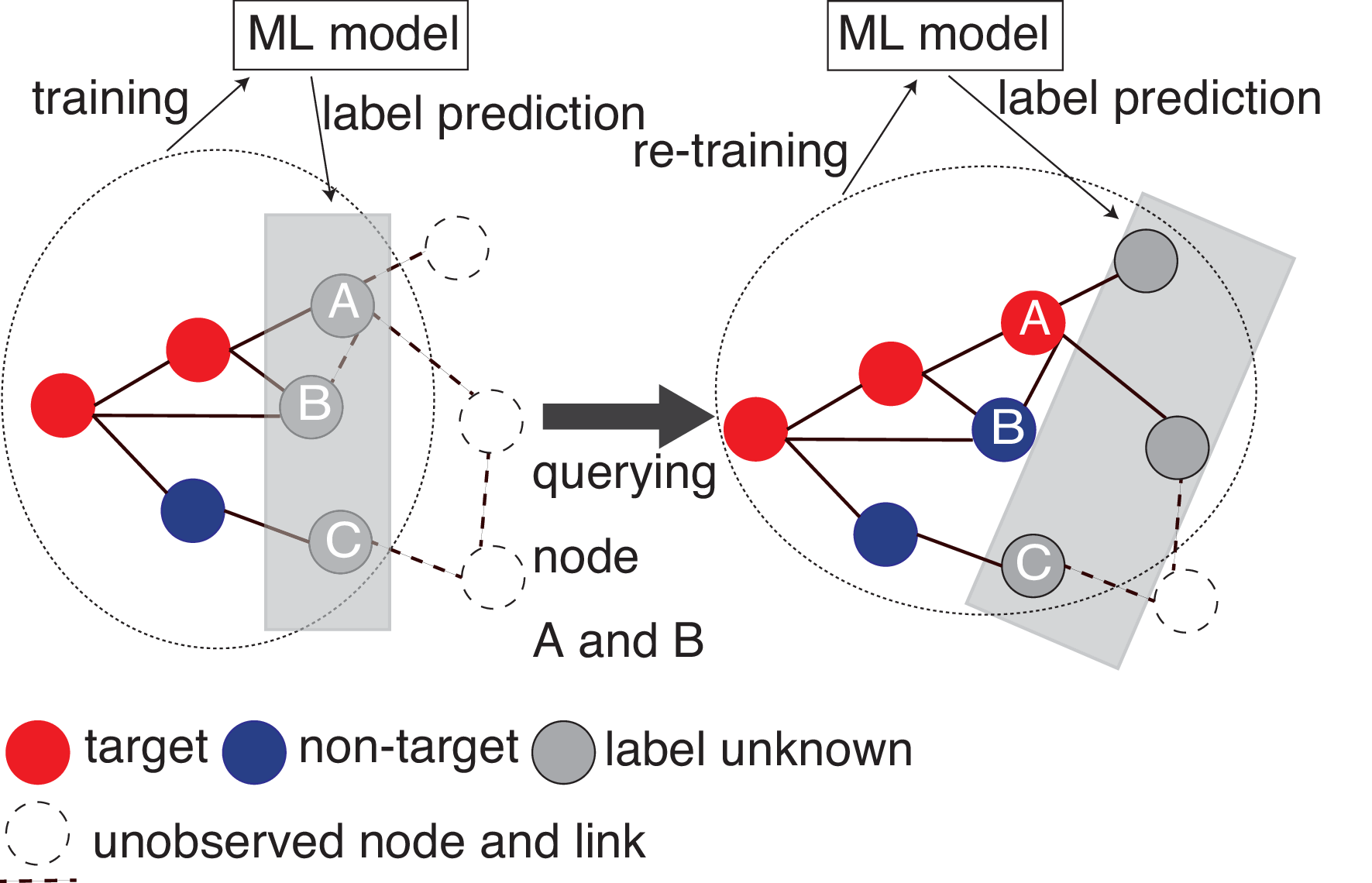}  
\caption{Overview of problem of discovering hidden target nodes. The red and blue nodes have been queried, and their labels are already known. The gray nodes A, B, and C are adjacent to the queried nodes but have not yet been queried, so their labels are unknown. We train a machine learning (ML) model to predict the unknown node labels by using the ones already known and the observed network topology, and we use the trained ML model to determine which nodes to query.  Here, nodes A and B are queried, and their labels are revealed. We then retrain the ML model using the newly observed topology and node labels, then we determine the nodes to query in the next round. By repeating this procedure, we aim to discover as many target nodes as possible with a given query budget.}
  \label{fig:overview}
\end{figure}

We aim to solve the problem of hidden-node discovery through graph
exploration based on ML.
In our approach, we assume that by querying a node in a graph, we gain access to its true label and its adjacent nodes. 
Note that the true labels of the adjacent nodes of the queried node remain unknown.
Based on the structure of the subgraph obtained via exploration and the label
information of the queried nodes, we then decide which nodes to query next. 
To do this, we could use a predictive model to determine whether a node is a target node, but unlike traditional node-label estimation problems where complete graph
structures are available, here we can only rely on incomplete subgraph
structures, which distinguishes this problem from typical node-label
estimation tasks~\cite{tsugawa22:icwsm,kim2023influencerrank,wang2017sybilscar,furutani2023interpreting,sun2020trustgcn}.
A challenge of this task lies in deriving node features suitable for node label estimation from incomplete structures.
Figure~\ref{fig:overview} illustrates the problem studied in this paper.

The specific research questions addressed in this paper are as follows.
\begin{itemize}
\item {\bf RQ1} How does the efficiency of hidden-node discovery differ between known and unknown topology of the target graph to be explored?
\item {\bf RQ2} How effective are node embeddings~\cite{Rossi18:TKDE} in discovering target hidden nodes?
\item {\bf RQ3} Can different types of target nodes be discovered using a unified framework? If so, what specific graph exploration strategies are required?
\end{itemize}
To address these questions, we extensively evaluate the efficiency of graph exploration strategies for target-node discovery on various social graphs, varying the prior knowledge about graph topology, and features used for target-node prediction.

\section{Literature Review}
\label{sec:related}

Also known as the label estimation problem, the problem of identifying target nodes in known graphs is a central task in graph mining and has engaged numerous researchers~\cite{tsugawa22:icwsm,kim2023influencerrank,wang2017sybilscar,mendoza2020bots,fazil2021deepsbd,feng2021botrgcn,ferrara2016rise,wang2020heterogeneous,darwish2020unsupervised}. The label
estimation problem includes the task of identifying influencers in
social graphs~\cite{tsugawa22:icwsm,kim2023influencerrank}, users with specific political
ideologies~\cite{wang2020heterogeneous,darwish2020unsupervised}, and social bots~\cite{mendoza2020bots,fazil2021deepsbd,feng2021botrgcn}. In such studies, supervised ML
is typically used to identify target
nodes, with models trained to predict whether an unlabeled node is
a target node based on the labeled nodes and their feature
vectors~\cite{tsugawa22:icwsm,kim2023influencerrank,mendoza2020bots,fazil2021deepsbd,feng2021botrgcn}. The goal
in node-label estimation problems is to construct highly accurate models.

Another fundamental problem in graph mining is graph exploration (also known as graph sampling)~\cite{chiericetti2016sampling,Gjoka10:Walking,Maiya11:Benefits,avrachenkov2014pay}.
Because the social graphs of interest are typically huge and have limited accessibility, graph exploration is a topic that is attracting much research interest.
The objectives of graph exploration vary, including unbiased
estimation of graph characteristics~\cite{Gjoka10:Walking,kurant2011towards} and maximizing coverage
of the target graph~\cite{avrachenkov2014pay,soundarajan2016maxreach}. Traverse-based exploration methods such as breadth-first search~\cite{kurant2011towards}, random walks~\cite{da2007exploring}, RDS~\cite{heckathorn1997respondent,mccreesh2012evaluation}, and their variations~\cite{Xie18:Efficient,Gjoka10:Walking} are widely used graph exploration techniques.

The study of target-node discovery problems in unknown graphs, which
combines node-label estimation and graph exploration, represents a pioneering
and challenging new problem. 
While this topic has been the subject of fewer studies to date than have graph exploration and label estimation individually, several pioneering studies have been conducted~\cite{murai2018selective,wang2020heterogeneous,morales2021selective}. Murai et~al.~\shortcite{murai2018selective} addressed a similar problem to the present target-node discovery in unknown graphs; they constructed classifiers to predict node attributes from the partially revealed graph structure through exploration and proposed the D$^3$TS strategy, which determines the nodes for exploration based on the predictions of the classifiers, and our study builds on theirs.
Furthermore, Morales et~al.~\shortcite{morales2021selective} proposed a target-node discovery method using deep reinforcement learning, and Wang et~al.~\shortcite{wang2020heterogeneous} introduced a method for discovering target nodes in heterogeneous networks composed of multiple types of nodes and links. However, these previous studies of target-node discovery in unknown graphs assumed either implicitly or explicitly that the target nodes are close to each other in the graph.
Previous studies have examined target-node discovery in unknown graphs through tasks such as identifying nodes belonging to specific categories in networks like research papers or Wikipedia articles, discovering nodes belonging to particular communities in online social networks or researcher networks, and finding donors of successful projects in crowdfunding platforms~\cite{murai2018selective}. In these cases, the correlation between adjacent nodes could be leveraged, as the target nodes are often in close proximity within the network. In contrast, our study extends the target-node discovery problem to more complex tasks, such as peripheral-node detection and influencer identification, where such proximity assumptions may not hold.

Following on from these previous studies, our main contributions are as follows.
\begin{itemize}
 \item     We address three new, challenging hidden-node discovery problems in unknown graphs that can be approached using the target-node discovery framework. As mentioned above, previous research has assumed that target nodes are similar and located close to each other on the graph~\cite{murai2018selective,morales2021selective}. In such problem settings, the network correlation--that ``the neighbors of a target node are likely to also be target nodes''--can be leveraged to discover target nodes even in unknown graphs. However, this study tackles various types of hidden-node discovery tasks, including more challenging problems where this correlation does not hold. We aim to clarify the effectiveness of the target-node discovery framework proposed by Murai et al.~\shortcite{murai2018selective} in addressing these different types of hidden-node discovery tasks.

 \item We extend the framework of Murai et~al.~\shortcite{murai2018selective} by incorporating node embeddings~\cite{Rossi18:TKDE} to obtain node features for constructing classifiers.
 While Murai et~al.~\shortcite{murai2018selective} relied on basic graph features such as node degree and the number of adjacent target nodes, our study demonstrates the effectiveness of node embeddings---which have gained attention for their efficacy in label estimation problems---in target-node discovery tasks.

\end{itemize}

\section{Problem Formulation}
\label{sec:problem}

\subsection{Hidden-Node Discovery Problem}

We denote the graph under exploration as $G=(V,E)$. Each node $v \in V$ is labeled with an attribute label $l(v)$, and we assume that $l(v)$ indicates whether node $v$ is a target node.

Initially, the structure of the graph $G$ is unknown, but a subgraph $G_{0}$ of $G$ is provided; $G_{0}$ is obtained by querying $m_0$ nodes in $G$.
By querying a node $v$, we obtain its label $l(v)$ and the set of adjacent nodes $\mathcal{N}(v)$. If $G$ is a directed graph, then the adjacent nodes of $v$ include nodes adjacent to $v$ via either incoming or outgoing links. The total number of nodes and the node set in graph $G$ are unknown, and only the nodes contained in the initial subgraph $G_{0}$ are known.

\begin{figure}[t]
  \centering 
\includegraphics[clip,width=.8\columnwidth]{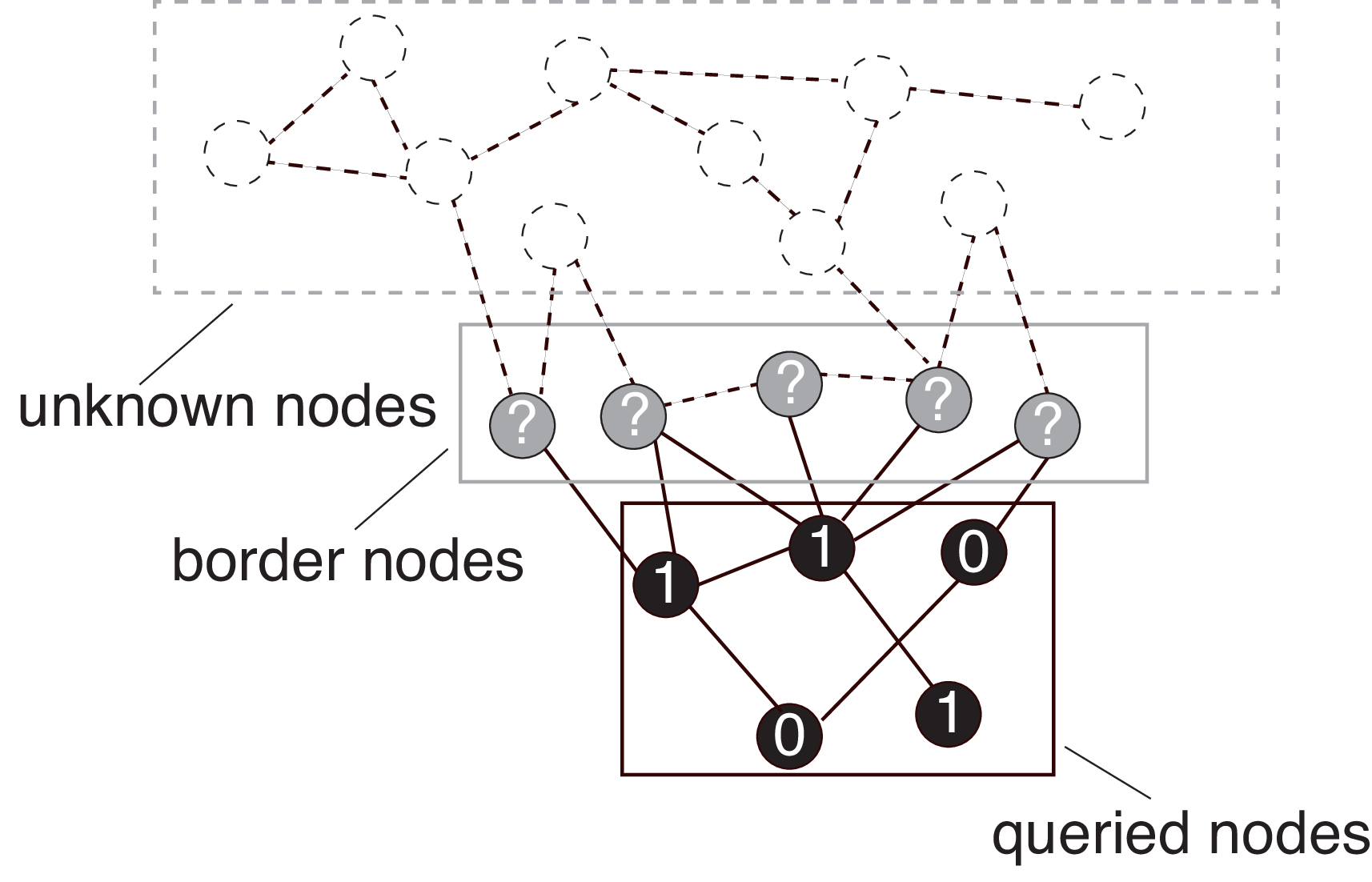}  
\caption{Graph exploration process. The solid black nodes are queried nodes, and their labels (1 or 0) are already known. The solid gray nodes are border nodes that are adjacent to queried nodes but have not yet been queried, and therefore their labels are still unknown. The dashed nodes and links are unknown ones. The subgraph composed of solid nodes and links is already known. In each round, we select the nodes to be queried from the border nodes by using the already known subgraph and the node labels of queried nodes.}
  \label{fig:problem}
\end{figure}

The exploration is conducted in $R$ rounds, with the number of queries
in each round denoted as $m_1,m_2,\ldots ,m_R$, and the total number of
queries as $M=\sum_{k=1}^R m_k $. In round $k$, the partial subgraph
$G_{k-1}$ obtained from previous rounds of exploration and the set of
labels of queried nodes are available. 
Note that $G_{k-1}$ contains both {\em queried nodes} and {\em border
nodes} that have not been queried in previous rounds.
We denote the set of queried nodes in round $k$ as $Q_k$, and the set of
border nodes in round $k$ as $B_k$.
For each node $v \in Q_k$, $l(v)$ is available, but for each node $u \in B_k$, $l(u)$ is unknown.
Using the information from
$G_{k-1}$ and the labels of queried nodes $Q_{k-1}$, $m_k$ nodes to be queried are
selected from border nodes $B_{k-1}$. 
 By querying $m_k$ nodes,
the graph $G_k$ is obtained. Using the information from $G_k$ and the
labels of queried nodes $Q_k$, $m_{k+1}$ nodes to be queried in the next round $k+1$ are further determined.
Figure~\ref{fig:problem} illustrates the graph exploration process.

In the present problem of discovering hidden target nodes, the
objective is to maximize the number of such nodes discovered after $R$
rounds (i.e., after $M$ nodes are queried). The hidden-node discovery
algorithm determines the nodes to be queried from the border nodes in each round to maximize
the number of discovered target nodes.
Note that each node's attribute label \(l(v)\) indicates whether the
node \(v\) is a hidden target node or not.

\subsection{Definitions of Target Node Labels}

This paper examines three different types of hidden-node discovery tasks: discovering Sybil nodes, peripheral nodes, and influencers. Sybil nodes tend to cluster together on the graph, while peripheral nodes and influencers are more widely dispersed. Furthermore, peripheral nodes have few connections within the graph, whereas influencers are highly connected. We tackle the challenge of discovering such various types of hidden nodes. The specific criteria for labeling target nodes in each of these tasks are detailed below.

\paragraph{Discovery of Sybil Nodes}

Sybil nodes are malicious nodes within social graphs, intent on spread fake news or manipulating public opinion~\cite{wang2017sybilscar,furutani2023interpreting,jia2017random,sun2020trustgcn}, and the Sybil-node discovery task aims to discover hidden malicious nodes on
a social graph.
To obtain the ground-truth labels of Sybil nodes on social graphs, we
adopt the standard setup for Sybil detection
studies~\cite{jia2017random,gong2014sybilbelief,gao2018sybilfuse}.
First, we duplicate a graph dataset $G_{\rm B}=(V_{\rm B},E_{\rm B})$, which represents the normal region, to create the Sybil region $G_{\rm S}=(V_{\rm S},E_{\rm S})$. The two graphs, $G_{\rm S}$ (Sybil region) and $G_{\rm B}$ (normal region), are structurally identical.
All nodes within $G_{\rm S}$ are designated as Sybil
nodes, labeled as $l(v)=1 \, (v \in V_{\rm S})$, while all nodes within
$G_{\rm B}$ are considered normal nodes, labeled as $l(v)=0 \, (v \in
V_{\rm B})$. We then randomly select $L$ pairs of Sybil nodes
from $G_{\rm S}$ and normal nodes from $G_{\rm B}$ and generate links
between these paired nodes. This process generates the entire graph to
be explored, denoted as $G$. This simulation mirrors the behavior of Sybil nodes collaborating and linking with each other, and the action of Sybil nodes generating links with normal nodes to obfuscate the distinction between them~\cite{jia2017random}. The difficulty of Sybil-node discovery increases as the number of links $L$ connecting Sybil nodes and normal nodes increases. In this paper, we set $L=80,000$ for all networks, corresponding to a relatively challenging setting for Sybil-node detection as observed previously~\cite{jia2017random}.

\paragraph{Discovery of Peripheral Nodes}

The peripheral-node discovery task is motivated by applications that require identifying individuals who are hard to reach because of sparse social connections (e.g., those who are in need of assistance but are challenging for administrative support organizations to locate, those who are experiencing poverty, or those who belong to marginalized social groups). In peripheral-node discovery, the target nodes are defined based on the coreness of each node in the target graph, using the $k$-core index~\cite{Dorogovtsev06:k}. Many social networks exhibit a core--periphery structure, comprising a core of nodes interconnected with numerous links and peripheral nodes connected to the core with few links~\cite{borgatti2000models}. By employing coreness, we can distinguish between peripheral nodes and those belonging to the core~\cite{kitsak2010identification}. In this paper, we compute the coreness of each node and designate the bottom 10\% of nodes in terms of coreness as peripheral nodes, that is, the target nodes. Thus, we assign $l(v)=1$ to the nodes that are in the bottom 10\% in terms of coreness ranking, and we assign $l(v)=0$ to the other nodes.

\paragraph{Influencer Discovery}

As defined in the present context, influencers are social media users who are capable of disseminating information to a wide audience. Following Tsugawa and Watabe~\shortcite{tsugawa22:icwsm}, we address the task of discovering two types of influencers: source spreaders and brokers. Source-spreader influencers can broadly propagate their own posts to numerous users, while broker influencers can disseminate the posts of others to a large audience~\cite{tsugawa22:icwsm}. Using the same Twitter Japan dataset as that referenced by Tsugawa and Watabe~\shortcite{tsugawa22:icwsm}, we compute source-spreader scores and broker scores for each node in the Twitter social graph, then we identify the top 10\% of nodes based on these scores as influential source spreaders and influential brokers. The source spreader score ${\mathcal S}_u$ and the broker score ${\mathcal B}_u$ of node $u$ are defined as ${\mathcal S}_u=\left|\bigcup_{c \in C_u} U_c \right|$, and ${\mathcal B}_u= \left|\bigcup_{c \in D} R_c^u \right|$, where $C_u$ is a set of cascades initiated by node $u$, $U_c$ is the set of users who retweet the original tweet of cascade $c$, $D=\{c\}$ is a set of diffusion cascades among the social media users in the dataset, and $R_{c}^u$ is the set of users who retweet in cascade $c$ after user $u$ retweets in cascade $c$.
Specifically, for the influential source spreader (resp.\ broker) discovery task, nodes with top-10\% influential source spreader (resp.\ broker) scores are labeled as $l(v)=1$, while the other nodes are labeled as $l(v)=0$. Influencers typically occupy the core of the network~\cite{kitsak2010identification}, distinguishing them from peripheral nodes.

\section{Methodology}
\label{sec:method}

\subsection{Datasets}

In this paper, we use the social graphs of Facebook~\cite{Leskovec12:learning}, Enron~\cite{leskovec2009community}, and
Epinion~\cite{richardson2003trust} for both the Sybil-node and peripheral-node discovery
tasks; these are publicly available datasets that were used in previous research on Sybil-node detection
within known graphs~\cite{jia2017random}. However, for the task of
influencer discovery, we use the social graph of the Twitter Japan dataset as
used in previous studies on influencer
identification~\cite{tsugawa22:icwsm}. Table~\ref{tab:data} presents an overview of
the social graphs.
Note that the datasets used in this paper are anonymized, and the hidden-node discovery tasks are not intended to expose confidential personal information.

\begin{table}[tb]
\centering

 \resizebox{.99\columnwidth}{!}{ \begin{tabular}{l|r|r|r|r}
\toprule
  & Facebook&
	  Enron&
	      Epinion & Twitter\\ \hline
Num. nodes  & 4,039& 33,696& 75,879 &351,759\\ 
Num. links & 88,234& 183,831&508,837 &  29,100,618\\
\bottomrule
 \end{tabular}}
\caption{Basic statistics of studied social graphs}
\label{tab:data}

\end{table}

\subsection{Strategies for Discovering Hidden Target Nodes}
\label{sec:algorithm}

In each exploration round, we determine the nodes to query by using an
ML-based approach. Specifically, in the \(k\)-th
exploration round, we estimate the probability of each node \(v \in
B_{k-1}\) being a target node based on the information available for the
graph \(G_{k-1} = (V_{k-1}, E_{k-1})\) in round \(k\). We use the
features vector \(\vb{f}_{k-1}(v)\) derived from \(G_{k-1}\), along with
the labels of already queried nodes \(Q_{k-1}\), to train a model
\(\mathcal{M}_{k-1}\) that predicts the probability of \(v\) being a
target node. The features of the queried nodes \(\{\vb{f}(v) \mid v \in
Q_{k-1}\}\) and their labels \(\{l_v \mid v \in Q_{k-1}\}\) serve as the
training data. The model \(\mathcal{M}_{k-1}\) takes the feature vector
\(\vb{f}_{k-1}(v)\) ($v \in B_{k-1}$) as input and outputs the
probability of \(v\) being a target node. 
We calculate the probability of being a target node for all border nodes $B_{k-1}$, then we select \(m_k\) nodes in descending order of this probability as the nodes to query in round \(k\).

The features of nodes for the prediction models are the basic features
used by Murai et~al.~\shortcite{murai2018selective}, 
and node embedding features.  The basic features are defined for each individual node $v$ based on its own information and that of its neighboring nodes. These features include: the number of adjacent target nodes to node $v$; the ratio of target nodes among all adjacent nodes of $v$; the number and ratio of triangles formed by target nodes and $v$; the number and ratio of triangles formed by non-target nodes and $v$; the number of triangles formed by adjacent nodes and $v$; and the number of target nodes reachable within two hops from $v$.

The node embedding features are obtained from the
DeepGL algorithm~\cite{Rossi18:TKDE}.  
In this paper, we must obtain consistent node embedding features in
an evolving graph because the topological structure of the
target graph changes in each round.
Therefore, we use an inductive node embedding algorithm, DeepGL~\cite{Rossi18:TKDE}, which is applicable to evolving graphs. By leveraging DeepGL, we can obtain consistent embedding vectors across graph $G_i$ in a certain round $i$ and graph $G_j$ in different rounds $j$.  The effectiveness of DeepGL embeddings has been demonstrated for the problem of influencer identification in known graphs~\cite{tsugawa22:icwsm}. Among various inductive node embedding methods~\cite{Rossi18:TKDE,hamilton2017inductive}, we select DeepGL~\cite{Rossi18:TKDE}.

Below, we briefly introduce DeepGL (see~\cite{Rossi18:TKDE} for more details). DeepGL utilizes the base features $\vb{x}$ of each node and learns the vector representation of a node based on its own base features and those of its neighbors by applying a relational function $f$. A relational function combines relational feature operators that can be applied to a base feature, including mean, sum, and max. In DeepGL, each dimension of the learned representation vector for a node is defined as a combination of its base features and relational operators.
DeepGL learns complex node features from the network in an unsupervised manner by combining base features with relational functions. The base features used in this paper are the same basic features described by Murai et al.~\shortcite{murai2018selective}, along with the node label $l(v)$. Note that for border nodes, $l(v)$ is unknown, and we assign $l(v)=-1$ to these nodes.
The model parameters are provided in Tab.~\ref{tab:DeepGL}.

\begin{table}[tb]
 \centering

 \resizebox{.99\columnwidth}{!}{ \begin{tabular}{l|l}
\toprule
Base features & basic features used in~\cite{murai2018selective},\\ 
& node label $l(v)$ \\
Relational functions & sum, max, mean \\
$\lambda$ & 0.7\\
Ego network distance & 2 \\
Transform method & log binning \\
\bottomrule

\end{tabular}}
\caption{Parameter configurations for DeepGL.}
\label{tab:DeepGL}
\end{table}

To construct the target-node prediction model \(\mathcal{M}_{k}\) for each round $k$, we use LightGBM~\cite{ke2017lightgbm}, 
a widely used gradient boosting framework based on tree-based learning algorithms. LightGBM is particularly effective for handling large datasets with lower computational costs compared to other boosting methods, and it is well-suited for machine learning tasks including influencer identification~\cite{tsugawa22:icwsm}.
We use the default hyperparameters from the Python LightGBM package\footnote{\url{https://lightgbm.readthedocs.io/en/latest/index.html}}, and we compare the
exploration efficiency using two types of models: one using only basic
features (base) and another using DeepGL features as well as basic
features (DeepGL).

As baselines without ML, we use degree-based search
[maximum observed degree (MOD)]~\cite{murai2018selective,Maiya11:Benefits,avrachenkov2014pay} and target neighbor (TN) search~\cite{murai2018selective}. The MOD strategy
queries \(m_k\) nodes selected in descending order of node degree in
\(G_{k-1}\). The TN strategy queries \(m_k\) nodes selected in descending order of
the number of adjacent target nodes \(n(v) = |\{u \mid u \in
\mathcal{N}(v) \cap l(u)=1\}|\) in \(G_{k-1}\). These strategies were also 
used by Murai et~al.~\shortcite{murai2018selective}.

For comparison, we also employ the strategy of using LightGBM to determine
the nodes to be queried when the graph structure \(G\) is known. By
replacing \(G_{k-1}\) with \(G\), we construct a target-node prediction
model for the known graph structure, which is equivalent to solving the
target-node discovery problem as a node-label estimation problem on the
known graph. We consider the number of target nodes that are discoverable by this
method as being the upper bound of practical efficiency for the discovery of hidden target nodes.

\subsection{Parameter Settings}

Unless stated otherwise, in the following sections we use the
following parameter settings. We obtain the initial graph \(G_0\) via a random walk starting from a randomly selected node.
The number of initial queries \(m_0\) is 200 for the Facebook graph and
2000 for the other graphs.
The number of queries \(m_k\) remains constant across all rounds; we use $m_k=100$ for the Facebook graph and $m_k=1000$ for the other graphs.
The DeepGL model for obtaining node embeddings is constructed after completing the first half of the initial queries. 
 We retrain the target-node prediction model \(\mathcal{M}_{k-1}\) in
 each round \(k\) using the already queried nodes as training
 data. 
Also, we compute the features \(\vb{f}_{k-1}(v)\) for the border nodes $B_{k-1}$ in each round $k$. When computing the DeepGL
 embedding features, we apply the constructed DeepGL model to the graph
 \(G_{k-1}\) to obtain \(\vb{f}_{k-1}(v)\). 
Except for the Twitter dataset, we perform 10 trials of graph exploration while changing the initial graph $G_0$, and the subsequent results represent the average for these 10 trials; for the Twitter dataset, we use three trials.
All experiments were conducted on a workstation with an Intel Xeon E5-2620~v4 CPU and 128\,GB of memory.
The source codes used for the experiments is available at our GitHub repository\footnote{\url{https://github.com/s-tugawa/hidden_node_icwsm25/}}. 
We used the DeepGL implementation by~\cite{fujiwara2022network}.

\section{Results}
\label{sec:result}

\begin{figure*}[tb]
\centering
\subfloat[Facebook-Sybil]{\includegraphics[width=.66\columnwidth,height=3.4cm]{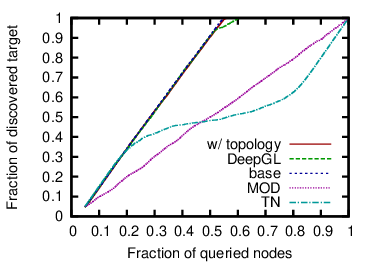}}
\subfloat[Enron-Sybil]{\includegraphics[width=.66\columnwidth,height=3.4cm]{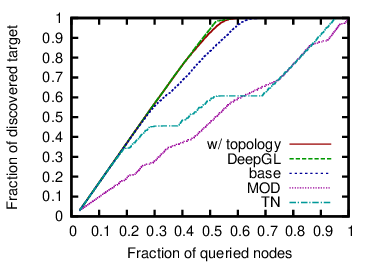}}
\subfloat[Epinion-Sybil]{\includegraphics[width=.66\columnwidth,height=3.4cm]{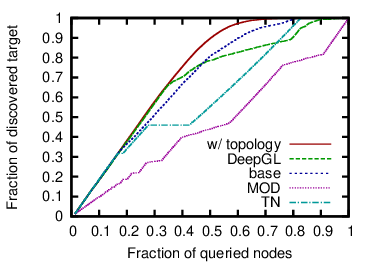}} 
\\
\subfloat[Facebook-Periphery]{\includegraphics[width=.66\columnwidth,height=3.4cm]{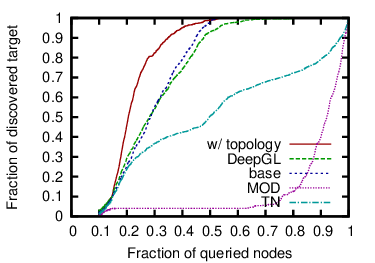}}
\subfloat[Enron-Periphery]{\includegraphics[width=.66\columnwidth,height=3.4cm]{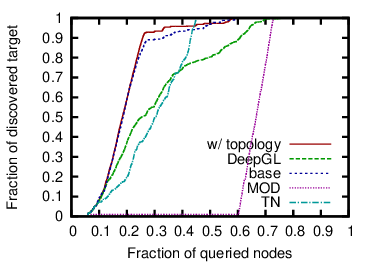}}
\subfloat[Epinion-Periphery]{\includegraphics[width=.66\columnwidth,height=3.4cm]{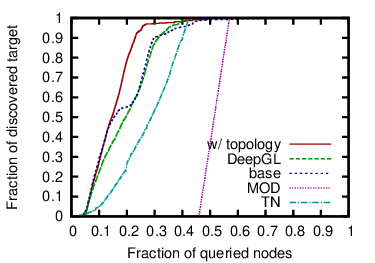}} \\
\subfloat[Twitter-Source]{\includegraphics[width=.66\columnwidth,height=3.4cm]{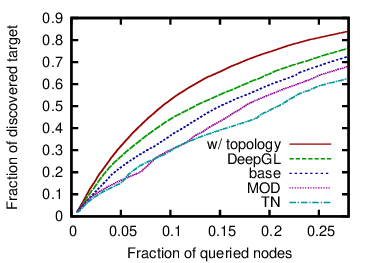}}
\subfloat[Twitter-Broker]{\includegraphics[width=.66\columnwidth,height=3.4cm]{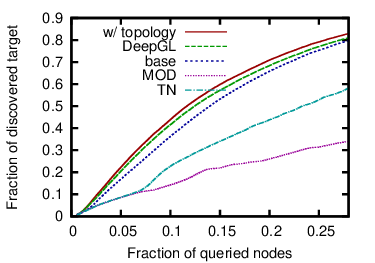}} 
    \caption{Fraction of discovered targets vs.\ fraction of queried nodes. 
The ML-based strategies (DeepGL and base) achieve efficiencies that are comparable with that of having complete knowledge about the target graph (w/~topology). While the DeepGL strategy is more efficient than the base strategy in influencer identification tasks (Twitter-Source and Twitter-Broker), it performs poorly in some other settings (e.g., Enron-Periphery and Epinion-Sybil).}
    \label{fig:coverage}
\end{figure*}

The relationships between the fraction of queried nodes and the fraction of discovered target nodes for each exploration strategy are shown in Fig.~\ref{fig:coverage}. Because of the substantial computational burden associated with exploring huge graphs, graph exploration was stopped upon querying 100,000 nodes (i.e., $R=100$ rounds) in the Twitter graph.
In Fig.~\ref{fig:coverage}, ``w/~topology'' represents the results obtained under the condition of known topological structure. Although experiments were conducted under this condition using both the DeepGL and base models, only the results from the more efficient one are presented: for Enron-Periphery and Epinion-Periphery, it is the results of the base model that are shown as w/~topology, while for the other cases it is the results from the DeepGL model that are shown as w/~topology.
Furthermore, ``DeepGL'' and ``base'' represent the results obtained when exploring under the condition of unknown topology using the DeepGL model and the base model, respectively.

\subsection{Known Topology vs.\ Unknown Topology}

First, we examine how the efficiency of hidden-node discovery differs between known and unknown topology of the exploration target ({\bf RQ1}).
Figure~\ref{fig:coverage} shows that the ML-based strategies
(DeepGL and base) achieve efficiencies that are comparable with that of having complete knowledge about the target graph (w/~topology).
Even in the case of identifying influential source spreaders on Twitter (where the difference between known and unknown topology is greatest), when 10\% of nodes have been queried, the DeepGL-based method has discovered over 80\% of the number of target nodes discovered when the entire graph structure is known (w/~topology). This suggests that
even when the complete graph structure is unknown, leveraging partially
observed graph structural features can enable efficient discovery of hidden target nodes. 

\subsection{Machine Learning vs.\ Heuristics}

Focusing on the differences among strategies, there is a considerable performance gap between the non-ML strategies (MOD and TN) and the ML-based strategies (DeepGL and base) in all settings, which suggests the difficulty of discovering hidden nodes. This difference is particularly pronounced when discovering many target nodes. For example, in the setting of Sybil-node discovery, the TN strategy performs comparably to the other strategies when discovering 20\% of Sybil nodes, but there is a considerable difference between the ML-based strategies and the TN strategy when discovering a higher percentage of Sybil nodes, such as over 90\%. This suggests that for efficient discovery of hidden target nodes in unknown graphs, exploiting partially observed graph structural features is crucial, and simple heuristics are insufficient, particularly when discovering many target nodes.

We anticipated that the hidden node discovery tasks in this study would be more challenging compared to the previous study~\cite{murai2018selective} where target nodes are close to each other in the graph. The low effectiveness of non-ML-based strategies supports this prediction. However, a surprising finding was that even under conditions where the graph structure was unknown, we were able to achieve exploration efficiency comparable to that of cases where the graph was fully known. We believe that, despite using features extracted solely from the local neighborhood of the target nodes, the combination of these features in a machine learning model enabled us to efficiently discover the hidden nodes.

\subsection{Using Node Embeddings is a Double-Edged Sword}

Next, we examine {\bf RQ2} (How effective are node embeddings in discovering target hidden nodes?).
Figure~\ref{fig:coverage} shows that in the setting of identifying Twitter influencers (source spreaders and brokers), using DeepGL embeddings contributes considerably to the improvement of efficiency in discovering influencers. Compared to the strategy of using only basic features (base), the strategy of using DeepGL embeddings (DeepGL) discovers 20\%--30\% more target nodes, and this indicates the usefulness of DeepGL embeddings for discovering influencers.

However, in the setting of peripheral-node discovery in Enron, it can be observed that using DeepGL embeddings worsens the efficiency of discovering peripheral nodes. In this setting, exploring the graph using the base model is much more effective than using the DeepGL embedding model.
Moreover, when over 40\% of nodes are queried, using the TN heuristic is more effective than using DeepGL embeddings. Similarly, in the Sybil-node discovery in Epinion, the model without embeddings demonstrates higher efficiency than the model using DeepGL embeddings.

These findings suggest that using DeepGL embeddings can improve the efficiency of discovery considerably for some types of hidden nodes while potentially worsening it for others, indicating a double-edged sword.

\begin{figure}[tb]
\centering
\includegraphics[width=.45\columnwidth]{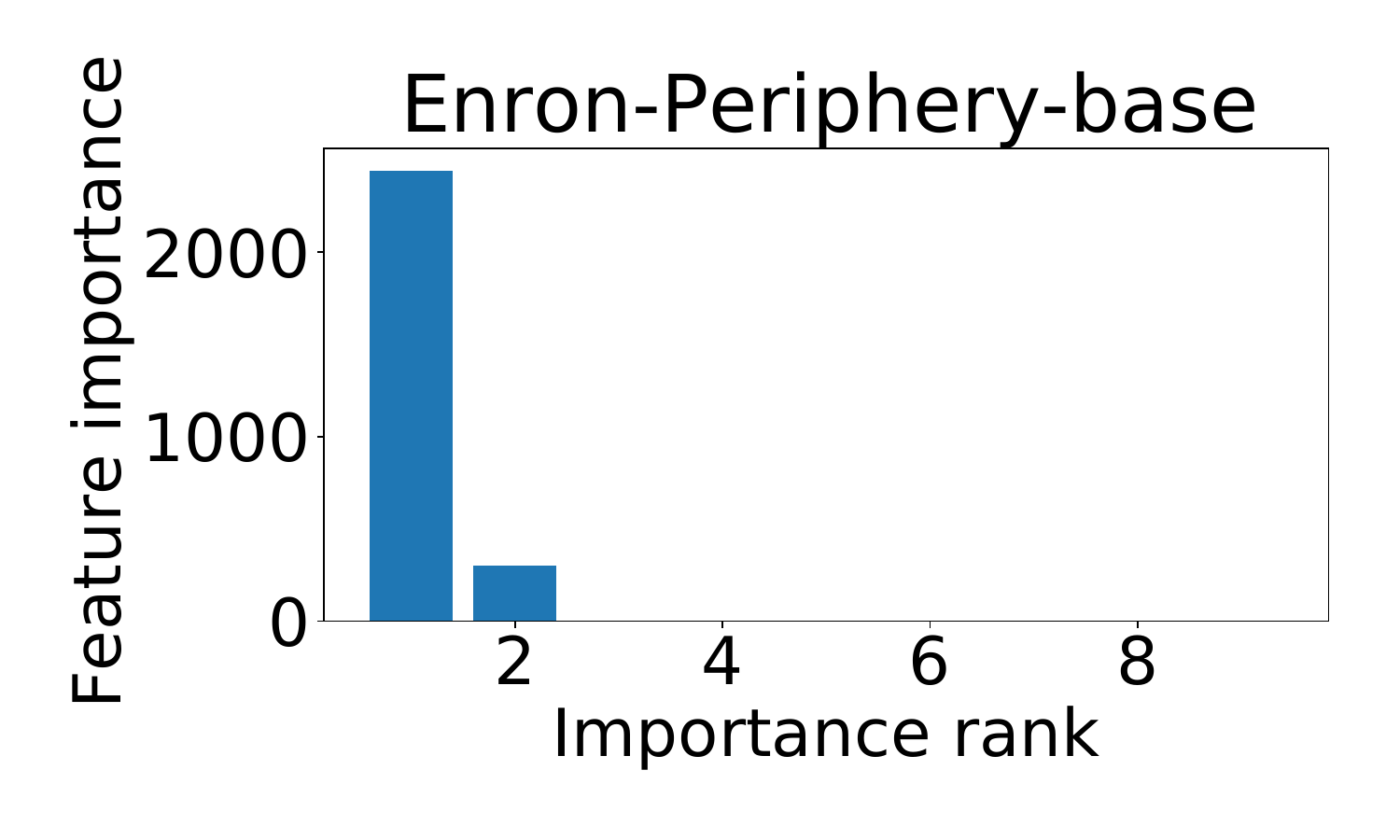}
\includegraphics[width=.45\columnwidth]{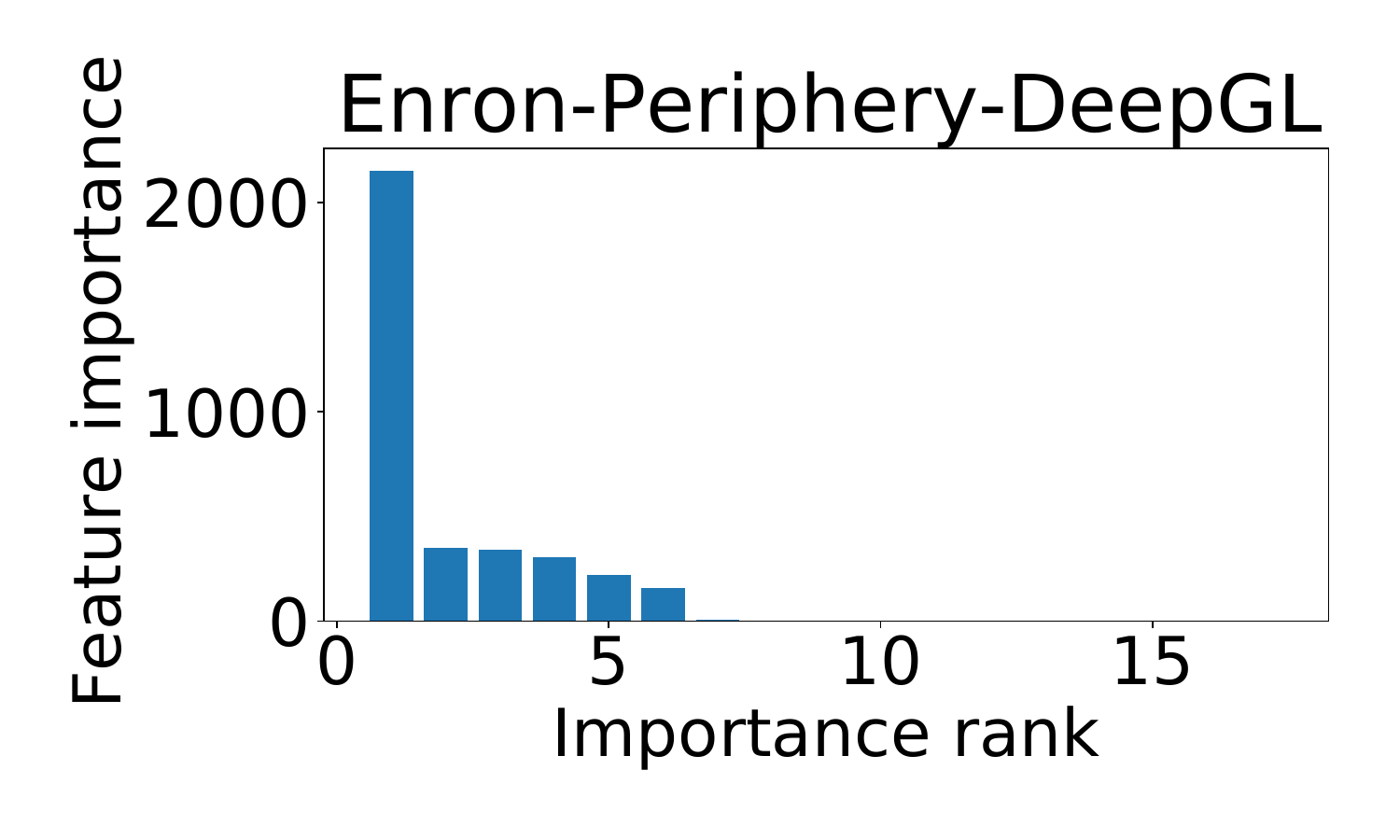} \\
\includegraphics[width=.45\columnwidth]{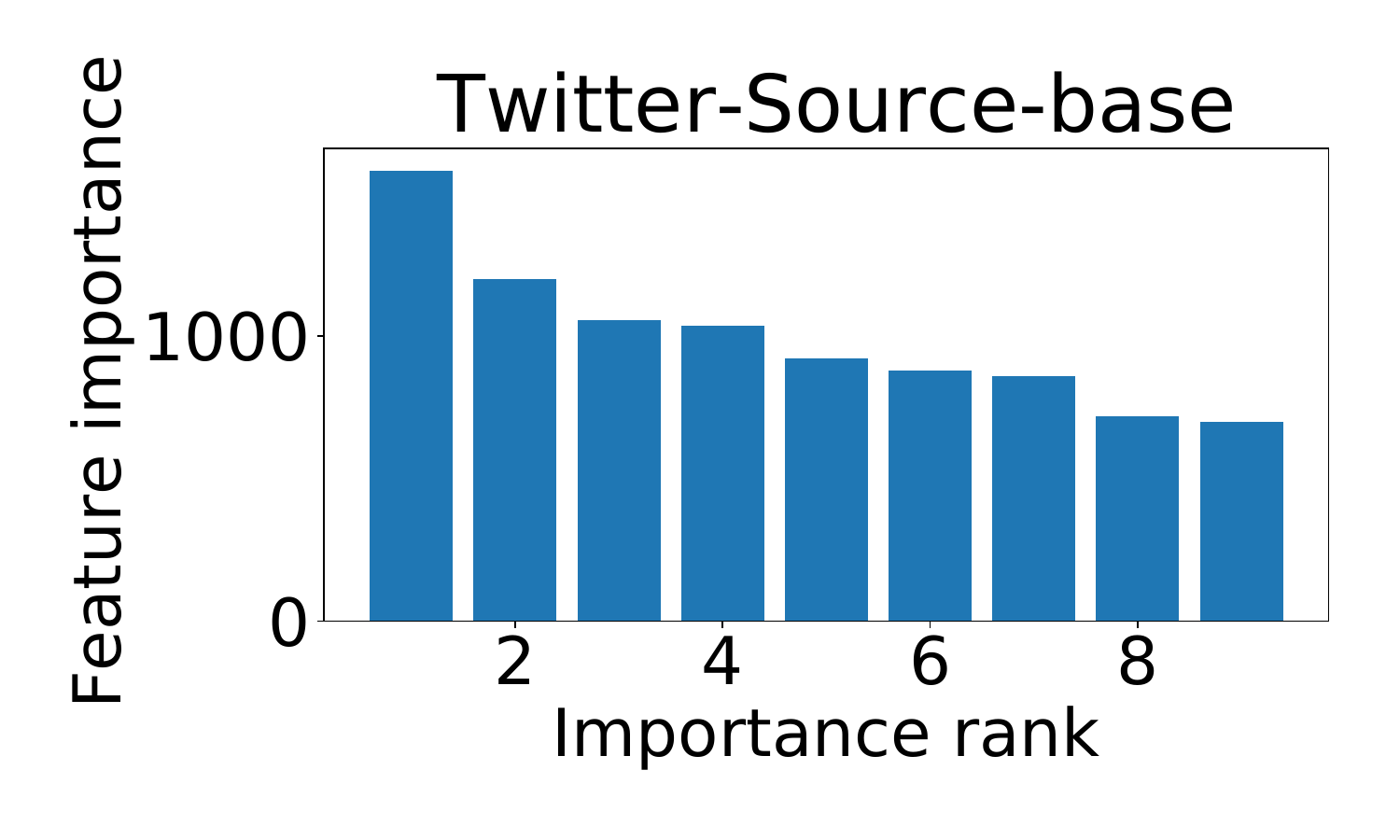}
\includegraphics[width=.45\columnwidth]{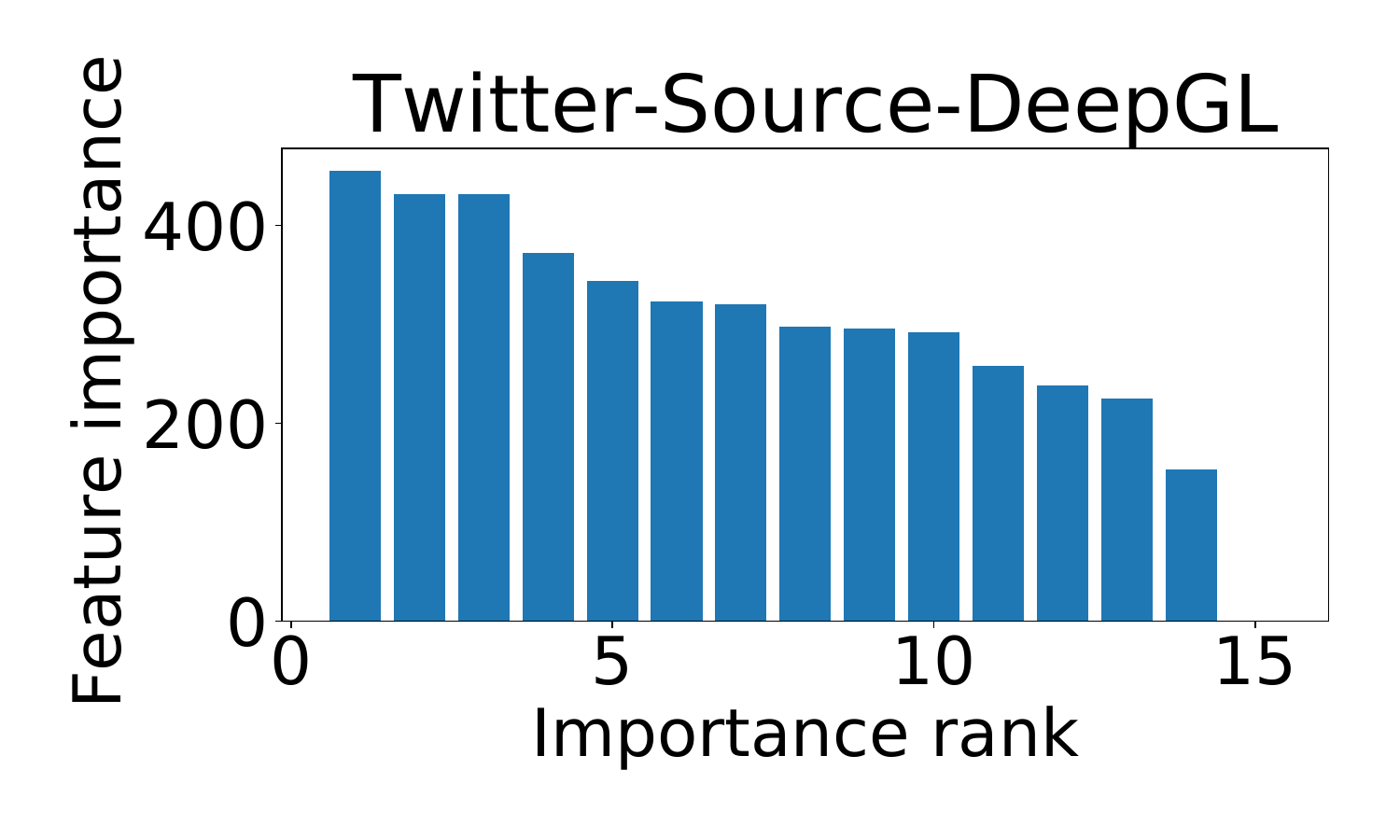}
    \caption{ Distributions of feature importance}
    \label{fig:importance}
\end{figure}

To further explore the effectiveness of embeddings, we investigated the feature importance of the base model and the DeepGL model at the final round (Fig.~\ref{fig:importance}). 
 Figure~\ref{fig:importance} shows the distribution of the Gini importance scores of features for the base and DeepGL models in two scenarios: Enron-Periphery, where embeddings had a particularly negative impact, and Twitter-Source, where embeddings were particularly effective.
From these results, we observe that in Enron-Periphery, only a few features contribute to the classification, whereas in Twitter-Source, many features contribute to a similar extent. This suggests that identifying influencers in Twitter-Source requires combining multiple features, while peripheral nodes in Enron-Periphery can be identified with only a few key features. Therefore, embeddings were effective in Twitter-Source, which required many features, but not in Enron-Periphery, where a small number of basic features were sufficient.
The decrease in exploration efficiency when embeddings were added to the base model in Enron-Periphery is likely due to overfitting caused by the inclusion of irrelevant embedding features.

\subsection{Ensemble of Multiple Classifiers is a Robust Strategy}

Now we address the final research question, namely, {\bf RQ3} (Can different types of target nodes be discovered using a unified framework? If so, what specific graph exploration strategies are required?). The previous findings indicate that while ML-based approaches are promising, the effectiveness of features varies across task settings. In other words, not all methods used in this study are efficient in discovering target nodes across all task settings.

To establish a unified strategy for hidden-node discovery irrespective of task settings, we investigate the application of bandit algorithms, which decide which option to choose from multiple options based on rewards obtained from selecting each option in the past.
Our previous experimental results suggest that effective strategies differ depending on the task at hand. Therefore, by using bandit algorithms to select the most effective strategy from multiple candidate strategies for the current task setting, it may be feasible to achieve efficient hidden-node discovery across different networks and task settings.
In the context of the problem studied in this paper, the multiple
options refer to deciding which model should be used for selecting
nodes to be queried.
Also, it is assumed that a reward is obtained when the queried
node turns out to be a target node. Essentially, rewards for selecting
the DeepGL model and the base model are recorded from past queries, and these rewards are used to stochastically favor the selection of models with higher rewards in future explorations. Among various bandit algorithms, we adopt D$^3$TS as used in the graph exploration study by Murai et~al.~\shortcite{murai2018selective}, using parameters identical to those in the literature. 
D$^3$TS is a multi-armed bandit algorithm based on Dynamic Thompson Sampling (DTS), and was proposed by Murai et al.\shortcite{murai2018selective} for graph exploration. 
In D$^3$TS, each arm $k$, corresponding to a classifier, is associated with two parameters: $\alpha_k$, representing the number of past successes, and $\beta_k$, representing the number of past failures. The choice of arm is determined by a random variable $r_k \sim \text{Beta}(\alpha_k, \beta_k)$ drawn from a Beta distribution. To ensure that the parameters remain normalized, the condition $\alpha_k + \beta_k < C$ is maintained for each arm. This prevents the algorithm from exclusively selecting the arm with the highest success rate and encourages exploration of other arms as well.
Murai et al.~\shortcite{murai2018selective} demonstrated that D$^3$TS is more effective compared to other bandit algorithms, such as $\epsilon$-greedy, which is why we adopt this method in our study.

\begin{figure}[tb]
\centering
\includegraphics[width=.99\columnwidth]{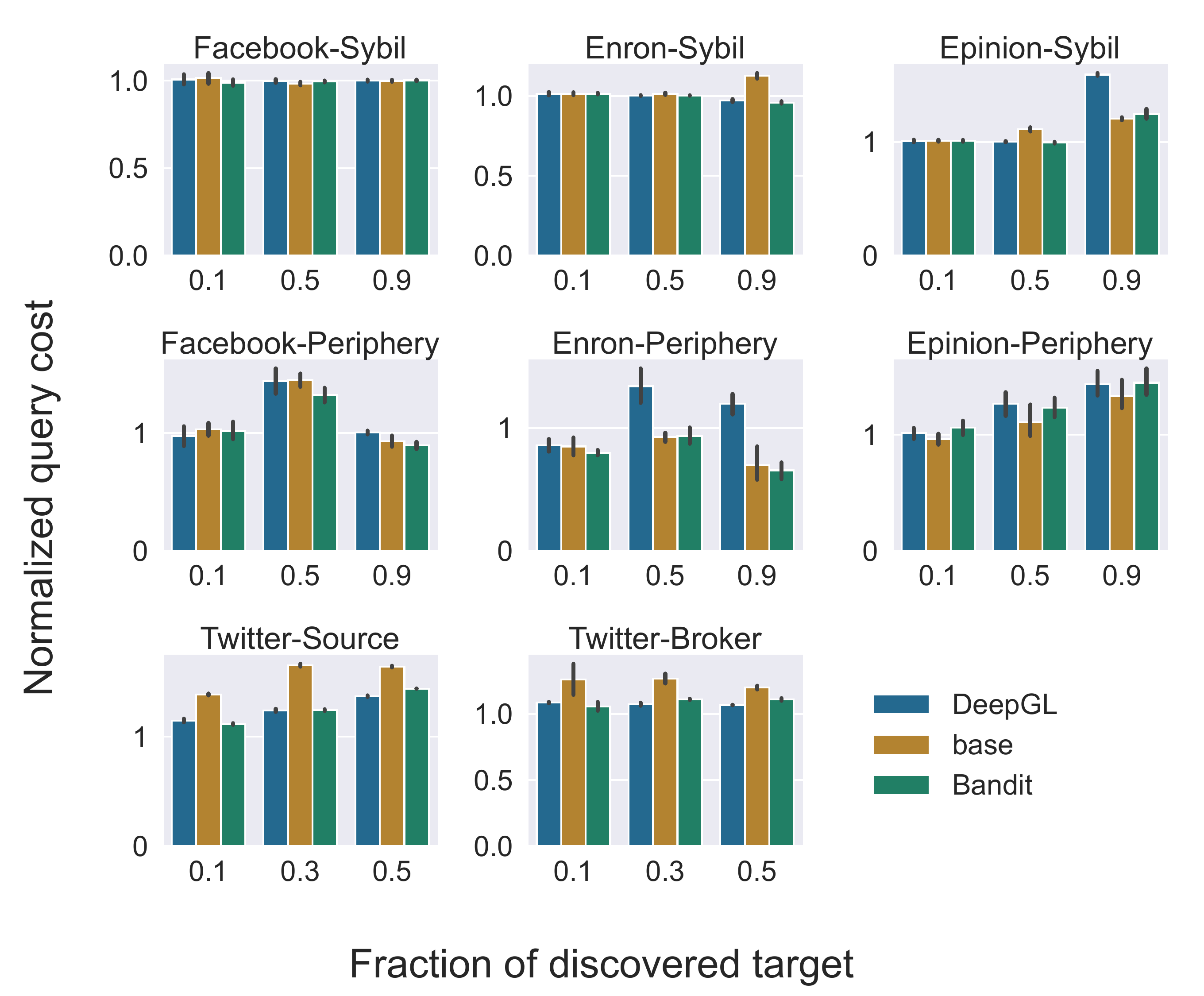}
    \caption{Comparison of normalized query cost among bandit, DeepGL, and base strategies. The bandit strategy consistently achieves a low query cost across most task settings.}
    \label{fig:bandit-time}
\end{figure}

Figure~\ref{fig:bandit-time} compares the efficiency of discovering hidden target nodes when using a bandit algorithm with the efficiency when only the DeepGL or base model is used. Here, as a measure of target-node discovery efficiency, we use the normalized query cost, which is defined as the number of queries required to discover a fraction $p$ of target nodes using the chosen strategy divided by the number of queries required to discover the same fraction of target nodes when complete knowledge about the topological structure is available. A lower normalized query cost indicates higher efficiency in discovering target nodes using that strategy.

Figure~\ref{fig:bandit-time} demonstrates that bandit algorithms consistently achieve high efficiency across all settings. For instance, in the Enron-Periphery setting, where the efficiency of the DeepGL strategy is very poor, the bandit strategy achieves an efficiency equivalent to that of the base strategy. Moreover, in the Twitter-Source and Twitter-Broker settings, where the efficiency of the base strategy is lower compared to the DeepGL strategy, the bandit strategy achieves an efficiency equal to or higher than that of the DeepGL strategy. While the effect of the bandit strategy is not particularly great in the Epinion-Periphery setting, overall it achieves efficiencies equal to or higher than those of the best-performing methods.
This suggests that using multiple classifiers with a bandit algorithm
can be effective in the absence of prior knowledge of target-node
features, and this strategy can be a unified framework for discovering
several types of target nodes on unknown social graphs.
In practice, it is often unclear which classification model will be most effective for hidden node discovery. Therefore, our finding that a bandit algorithm can learn which model to use during the graph exploration process is practically useful.

\subsection{Benefits of Model Retraining}

\begin{figure}[tb]
\centering
\includegraphics[width=.99\columnwidth]{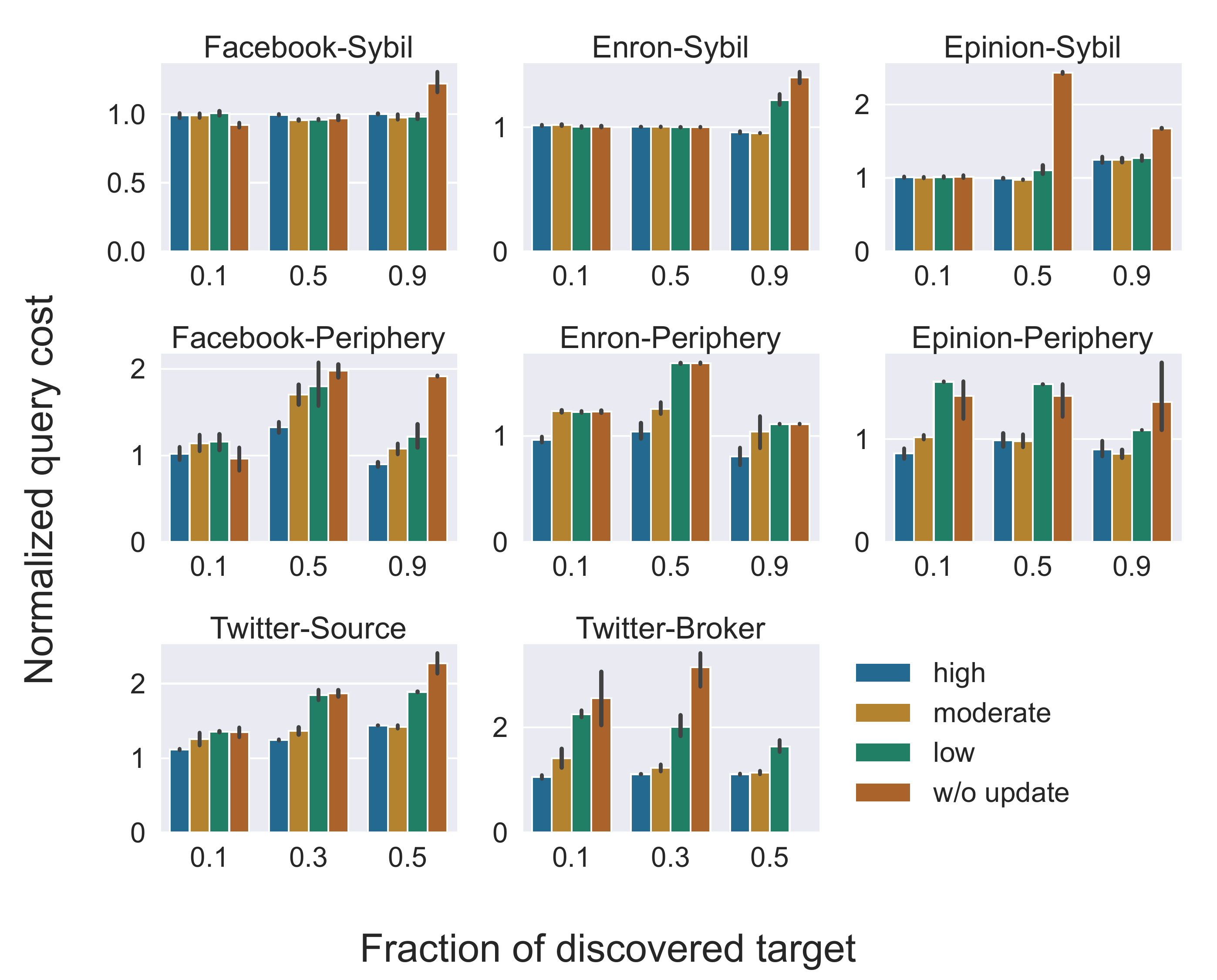}
    \caption{Comparison of normalized query costs for different model retraining frequencies. In the legend, high, moderate, and low correspond to update frequencies of $m=100$, $m=500$, and $m=1000$ respectively for Facebook data. For other cases, they correspond to $m=1000$, $m=5000$, and $m=30000$ respectively.  The query cost tends to be higher when the frequency of model retraining is lower.}
    \label{fig:batch-time}
\end{figure}

Finally, we verify the importance of model retraining.
Figure~\ref{fig:batch-time} compares the normalized query cost (as used in the previous subsection) when changing the frequency of model retraining.
In our problem settings, we retrain the target-node prediction models for each round.
Here, we investigate the normalized query cost while changing the number of queries $m$ in each round, with larger $m$ meaning that the model update intervals are longer (i.e., lower frequency of model retraining).
We also compare the scenario where the prediction models are trained on
the initial graph $G_0$, and that where model retraining is never performed (w/o~update).
As the graph exploration strategy, we used the bandit strategy.
Note that in the Twitter-Broker setting without model retraining, more
than half of the target nodes could not be discovered within 100,000
queries, which is why there are no ``w/o~update'' results for the scenario with a target-node proportion of 0.5.

These results show that model retraining is advantageous for efficient discovery of hidden target nodes. In most scenarios, the strategy of not updating the model (w/o~update) yields poor performance. Generally, a higher frequency of model updating is deemed favorable; however, in the context of Sybil discovery settings, extending the interval between model updates does not impact the performance appreciably. Nonetheless, it is evident that never retraining the initial model leads to a considerable decrease in efficiency.
These results suggest that retraining the model with the newly available data obtained during the graph exploration process is crucial for efficient hidden-node discovery.
Particularly in the early stages of exploration, when there is insufficient training data, the prediction accuracy of the model is not yet sufficient (see also Fig.~\ref{fig:prec} in the appendix). Retraining the model after a certain amount of exploration to improve prediction accuracy is considered useful for achieving efficient graph exploration.

\subsection{Comparison with Other ML Algorithms}

\begin{figure}[tb]
\centering

\includegraphics[width=.99\columnwidth]{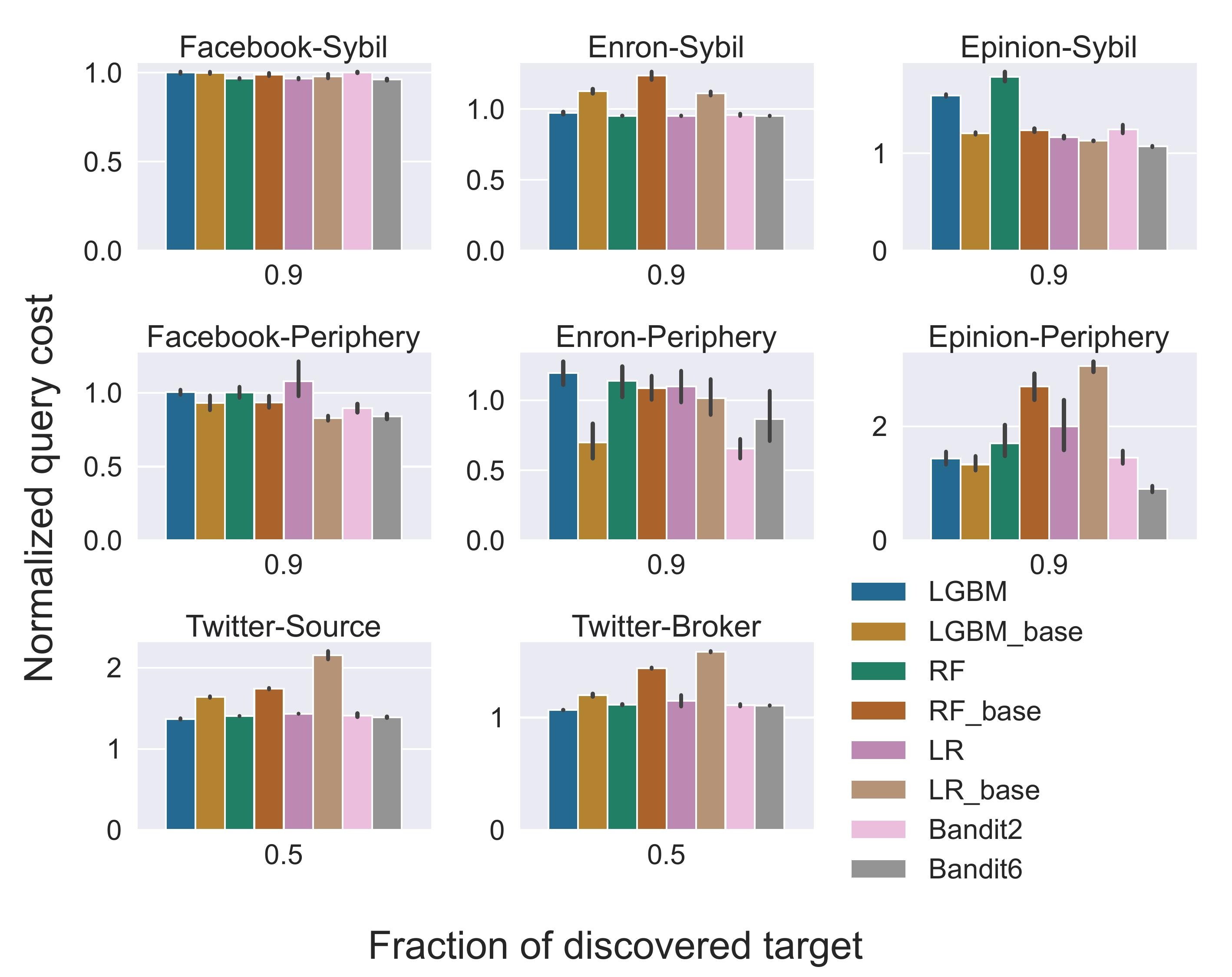}
    \caption{Comparison of normalized query cost among LR, RF, LGBM, and Bandit.The results labeled with base represent those using only basic features. Bandit2 refers to the combination of two models, LGBM with and without embeddings. Bandit6 combines six models: each classifier (LGBM, LR, RF) with and without embeddings.}
    \label{fig:model-time}
\end{figure}

In the previous experiments, LightGBM (LGBM) was used as the classifier, while the previous study on machine learning-based graph exploration~\cite{murai2018selective}, employed logistic regression (LR) and random forests (RF). Therefore, in this section, we compare the exploration efficiency of using LR and RF as classifiers against LGBM. For each classifier, we build two types of models: one using only basic features and the other incorporating DeepGL embeddings.
Previous research~\cite{murai2018selective} also proposed combining different classifiers with a bandit algorithm. Thus, we also evaluate the exploration efficiency of combining LGBM, LR, and RF using a bandit algorithm. The results are shown in Fig.~\ref{fig:model-time}. The results labeled with {\em base} represent those using only basic features. Bandit2 refers to the combination of two models—LGBM with basic features and LGBM with DeepGL embeddings—used in the experiments from previous sections. On the other hand, Bandit6 combines six models: each classifier (LGBM, LR, RF) with and without embeddings. Here, we also use the normalized query cost as the evaluation metric when discovering 50\% of the target nodes in the influencer discovery tasks and 90\% of the target nodes in the other tasks.

The results show the following observations: (1) The choice of classifier—whether LR, RF, or LGBM—has a relatively small impact on the results, while the use of embedding features has a significant effect. For example, in the influencer discovery task, models using embeddings achieved higher exploration efficiency compared to the base models, regardless of the classifier. Comparing LR, RF, and LGBM with the same features shows no substantial differences. (2) The difference between Bandit6, which combines all models, and Bandit2 is not significant. This suggests that while combining multiple classifiers generally has a positive impact on exploration efficiency, increasing the number of models does not lead to dramatic improvements. Since the exploration efficiency of each model is relatively similar, it is natural that the combination does not result in significant enhancements. Considering the training cost of each model, the benefit of combining many classifiers appears to be limited.

\section{Discussion}
\subsection{Implications}

The finding that comparable efficiencies can be achieved whether the topology structure of the exploration network is known or unknown was unexpected but offers practical advantages. 
In this paper, we trained a model to predict whether border nodes (whose connectivity is only partially known) are target nodes and then used it in graph exploration. Although this methodology has been used previously~\cite{murai2018selective,wang2020heterogeneous,morales2021selective}, those studies evaluated it only in cases in which the target nodes tend to be adjacent to each other, whereupon it is reasonable to predict the label of a border node based on its adjacency to target nodes. However, in the present settings of peripheral-node discovery and influencer discovery, such assumptions do not hold true.
 This is evident from the low performance of the TN heuristic, which prioritizes querying border nodes that are more adjacent to target nodes. 
Therefore, it is a non-trivial finding that target nodes can be predicted from border nodes with limited adjacency by fusing basic and embedding features. 
 This insight is expected to be valuable when estimating node characteristics in graphs with unknown structure.

The finding that using embedding features decreases the efficiency of
target-node discovery suggests the difficulty of discovering hidden nodes 
in unknown graphs. 
As discussed in the feature importance results in Fig.~\ref{fig:importance}, the drop in exploration efficiency when adding embedding features is a typical case of overfitting. In traditional machine learning tasks, if sufficient training data is available, LightGBM tends to disregard unhelpful features, making it unnecessary to be cautious about adding new features. However, in the graph exploration task tackled in this paper, the classification model is trained incrementally as data is gradually acquired. With insufficient training data, the presence of unhelpful features likely led to overfitting.
Furthermore, node embedding techniques are
fundamentally designed to obtain node features in known
graphs, and so they may be ineffective when seeking node features in incomplete graphs, as
in this study.
The present results suggest that caution is required when using node embeddings for mining tasks involving incomplete graphs~\cite{Wilder18:influence,tran2021community}.

The effectiveness of using a bandit algorithm across various settings aligns with the findings of Murai et~al.~\shortcite{murai2018selective}. Our results indicate that combining multiple ML models via bandit algorithms remains effective even when the implicit assumption of adjacency among target nodes is not met. Although the present study addressed three particular types of hidden-node discovery tasks, it suggests broader applicability of these findings to a wide range of tasks. On the other hand, increasing the number of classifiers used does not seem to provide significant additional benefits. This is likely because, when the same features are used, there are minimal differences in the strengths and weaknesses of the learned models, making the advantage of having more options relatively small. Considering the computational cost of model training, it is suggested that using a limited number of classifiers is a more reasonable approach.

\subsection{Limitations}

This study has some limitations, which we discuss below along with future research directions. First, the effectiveness of other features and classification models for target-node discovery tasks should be explored further. In particular, detailed examination of node embeddings suitable for hidden-node discovery problems is necessary. Although the present study used DeepGL~\cite{Rossi18:TKDE} as an inductive node embedding technique, its use did not consistently contribute to target-node discovery, indicating the need for research into node embedding methods suitable for hidden-node discovery in unknown graphs.

Second, to validate the practical effectiveness of our approach, we need a greater number of realistic datasets pertaining to hidden or hard-to-reach populations. Because social graph data on true hidden populations---such as Sybil nodes or HIV-infected individuals---are not publicly available, this study experimented with pseudo-labels of nodes on real social graphs, but to ascertain whether our method is genuinely useful for discovering hidden populations, we need field experiments that collect data from social graphs containing hidden populations. However, such field experiments involve not only technical challenges but also considerations of ethical, legal, and social issues regarding revealing individuals' hidden characteristics from social graph structures.

Third, the use of pre-trained models for target-node prediction should be discussed. While this study assumed no prior knowledge about target nodes in the initial state and constructed prediction models while exploring the graph, approaches such as using pre-trained models or combining pre-trained models with those constructed during exploration may be considered. Improving the efficiency of hidden-node discovery through such novel approaches remains an important task.

Finally, while we employ greedy strategies for hidden-node discovery, it is imperative to consider the exploration--exploitation tradeoff, as highlighted by Murai et~al.~\shortcite{murai2018selective}. We adopt the strategy of prioritizing nodes with a high probability of being the target node, based on the results of our predictive models. However, such greedy strategies are not universally optimal, and introducing some degree of randomness into the exploration process may enhance its efficiency. Moreover, from the perspective of model training, greedy exploration does not necessarily improve the predictive accuracy of the model. Prioritizing model training in the exploration process might enhance the efficiency of exploration. 
If random access to unknown nodes is permitted, integrating a random sampling strategy could be considered to balance the exploration-exploitation tradeoff. 
Investigating strategies for hidden-node discovery that account for such tradeoffs is a crucial future research endeavor.

\section{Conclusion}
\label{sec:conclusion}

In this paper, we addressed the problem of discovering hidden target nodes from unknown social networks, formulating three types of hidden-node discovery problems: Sybil-node discovery, peripheral-node discovery, and influencer discovery. We evaluated the effectiveness of ML-based graph exploration strategies for these problems, and the results showed that such strategies enable the discovery of hidden nodes with comparable efficiency to that in cases where the graph structure is known. Specifically, when using ML-based strategies, the query cost of discovering 10\% of the hidden nodes is at most only 1.2 times that in the case of known topology, and the query cost of discovering 90\% of the hidden nodes is at most only 1.4 times greater. Furthermore, we examined node features useful for discovering target nodes. While node embedding based on DeepGL~\cite{Rossi18:TKDE} proved highly effective for certain types of hidden nodes, its use sometimes degraded the efficiency of hidden-node discovery. On the other hand, by using a bandit algorithm~\cite{murai2018selective} to combine models with and without node embedding features, efficient hidden-node discovery was achieved across most settings, regardless of the type of hidden nodes.

\section{Acknowledgments}

 This work was supported by JSPS KAKENHI Grant No. JP22K11990.

\bibliographystyle{aaai25}
\bibliography{influence-short,sampling-short}

\section{Paper Checklist}

\begin{enumerate}

\item For most authors...
\begin{enumerate}
    \item  Would answering this research question advance science without violating social contracts, such as violating privacy norms, perpetuating unfair profiling, exacerbating the socio-economic divide, or implying disrespect to societies or cultures?
    \answerNo{No, because hidden node discovery strategies have the potential to unveil undisclosed personal attribute information, which may raise privacy concerns.  Please see Ethical Statement.} 
  \item Do your main claims in the abstract and introduction accurately reflect the paper's contributions and scope?
    \answerYes{Yes}
   \item Do you clarify how the proposed methodological approach is appropriate for the claims made? 
    \answerYes{Yes}
   \item Do you clarify what are possible artifacts in the data used, given population-specific distributions?
    \answerYes{Yes}
  \item Did you describe the limitations of your work?
    \answerYes{Yes}
  \item Did you discuss any potential negative societal impacts of your work?
    \answerYes{Yes, see Ethical Statement.}
      \item Did you discuss any potential misuse of your work?
    \answerYes{Yes, see Ethical Statement.}
    \item Did you describe steps taken to prevent or mitigate potential negative outcomes of the research, such as data and model documentation, data anonymization, responsible release, access control, and the reproducibility of findings?
    \answerYes{Yes, see Ethical Statement.}
  \item Have you read the ethics review guidelines and ensured that your paper conforms to them?
    \answerYes{Yes}
\end{enumerate}

\item Additionally, if your study involves hypotheses testing...
\begin{enumerate}
  \item Did you clearly state the assumptions underlying all theoretical results?
    \answerNA{NA}
  \item Have you provided justifications for all theoretical results?
    \answerNA{NA}
  \item Did you discuss competing hypotheses or theories that might challenge or complement your theoretical results?
    \answerNA{NA}
  \item Have you considered alternative mechanisms or explanations that might account for the same outcomes observed in your study?
    \answerNA{NA}
  \item Did you address potential biases or limitations in your theoretical framework?
    \answerNA{NA}
  \item Have you related your theoretical results to the existing literature in social science?
    \answerNA{NA}
  \item Did you discuss the implications of your theoretical results for policy, practice, or further research in the social science domain?
    \answerNA{NA}
\end{enumerate}

\item Additionally, if you are including theoretical proofs...
\begin{enumerate}
  \item Did you state the full set of assumptions of all theoretical results?
    \answerNA{NA}
	\item Did you include complete proofs of all theoretical results?
    \answerNA{NA}
\end{enumerate}

\item Additionally, if you ran machine learning experiments...
\begin{enumerate}
  \item Did you include the code, data, and instructions needed to reproduce the main experimental results (either in the supplemental material or as a URL)?
    \answerYes{Yes, we have released our code and data on GitHub repository.}
  \item Did you specify all the training details (e.g., data splits, hyperparameters, how they were chosen)?
    \answerYes{Yes}
     \item Did you report error bars (e.g., with respect to the random seed after running experiments multiple times)?
    \answerYes{Yes}
	\item Did you include the total amount of compute and the type of resources used (e.g., type of GPUs, internal cluster, or cloud provider)?
    \answerYes{Yes}
     \item Do you justify how the proposed evaluation is sufficient and appropriate to the claims made? 
    \answerYes{Yes}
     \item Do you discuss what is ``the cost`` of misclassification and fault (in)tolerance?
    \answerYes{Yes, see Ethical Statement.}
  
\end{enumerate}

\item Additionally, if you are using existing assets (e.g., code, data, models) or curating/releasing new assets, \textbf{without compromising anonymity}...
\begin{enumerate}
  \item If your work uses existing assets, did you cite the creators?
    \answerYes{Yes}
  \item Did you mention the license of the assets?
    \answerNo{No, because the datasets used in this paper have no explicit license statements.}
  \item Did you include any new assets in the supplemental material or as a URL?
    \answerYes{Yes, the source code will be available as a URL upon acceptance.}
  \item Did you discuss whether and how consent was obtained from people whose data you're using/curating?
    \answerNA{NA}
  \item Did you discuss whether the data you are using/curating contains personally identifiable information or offensive content?
    \answerYes{Yes}
\item If you are curating or releasing new datasets, did you discuss how you intend to make your datasets FAIR?
\answerNA{NA}
\item If you are curating or releasing new datasets, did you create a Datasheet for the Dataset? 
\answerNA{NA}
\end{enumerate}

\item Additionally, if you used crowdsourcing or conducted research with human subjects, \textbf{without compromising anonymity}...
\begin{enumerate}
  \item Did you include the full text of instructions given to participants and screenshots?
    \answerNA{NA}
  \item Did you describe any potential participant risks, with mentions of Institutional Review Board (IRB) approvals?
    \answerNA{NA}
  \item Did you include the estimated hourly wage paid to participants and the total amount spent on participant compensation?
    \answerNA{NA}
   \item Did you discuss how data is stored, shared, and deidentified?
   \answerNA{NA}
\end{enumerate}

\end{enumerate}

\section{Ethical Statement}

Hidden-node discovery strategies have the potential to reveal undisclosed personal attribute information, thus raising privacy concerns. Furthermore, incorrect estimation of node attribute information may introduce biases into decision-making processes. Therefore, it is essential to consider the associated ethical, legal, and social issues when implementing hidden-node discovery strategies. First, the social ties and label information obtained from social networks for model training should be collected based on clear and explicit consent from the individuals involved. Also, the constructed models should only be used for specific purposes such as social welfare or enhancing services on social media platforms, and indiscriminate sharing should be avoided. There is a risk that malicious third parties could exploit the framework proposed in this paper, leading to the involuntary exposure of personal attribute information. While such risks are inherent in advanced ML systems, addressing these risks is imperative when implementing hidden-node discovery strategies. Using authentication mechanisms during model usage and frameworks for privacy-preserving data mining can mitigate these risks, but further research is needed to assess their feasibility. In this study, we conducted model construction and evaluation using non-sensitive attribute information to verify the technical feasibility of hidden-node discovery. When using more-sensitive hidden labels in the future, it is crucial to consider the concerns discussed here.

\appendix

\section{Appendix}

\subsection{Comparison of Model Accuracy in Terms of Precision}

\begin{figure*}[tb]
\centering
\subfloat[Facebook-Sybil]{\includegraphics[width=.66\columnwidth]{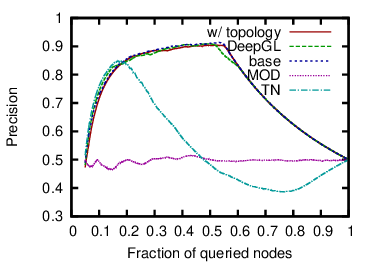}}
\subfloat[Enron-Sybil]{\includegraphics[width=.66\columnwidth]{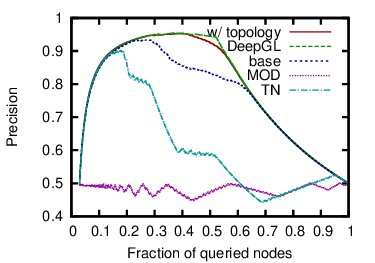}}
\subfloat[Epinion-Sybil]{\includegraphics[width=.66\columnwidth]{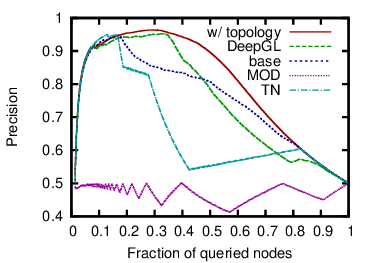}} 
\\
\subfloat[Facebook-Periphery]{\includegraphics[width=.66\columnwidth]{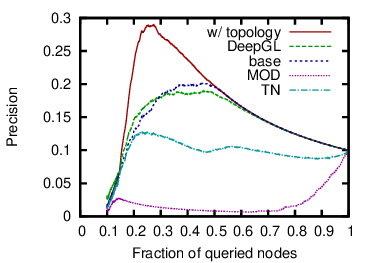}}
\subfloat[Enron-Periphery]{\includegraphics[width=.66\columnwidth]{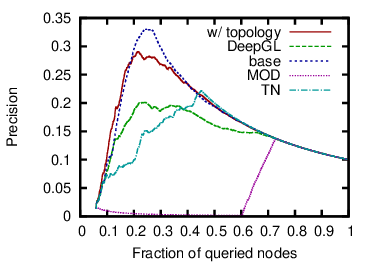}}
\subfloat[Epinion-Periphery]{\includegraphics[width=.66\columnwidth]{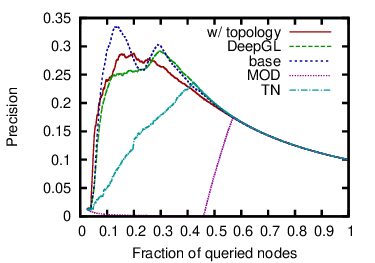}} \\
\subfloat[Twitter-Source]{\includegraphics[width=.66\columnwidth]{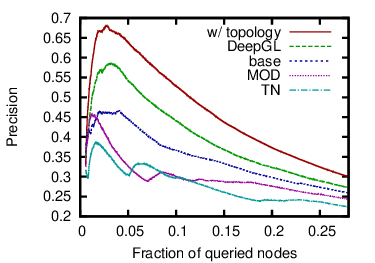}}
\subfloat[Twitter-Broker]{\includegraphics[width=.66\columnwidth]{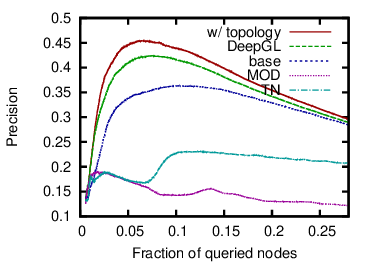}} 
    \caption{Precision vs.\ fraction of queried nodes. }
    \label{fig:prec}
\end{figure*}

As supplementary results for further analyzing exploration efficiency, Fig.~\ref{fig:prec} shows the relationship between the proportion of explored nodes and the precision, i.e., the proportion of target nodes among the explored nodes. This figure illustrates how the discovery accuracy of target nodes improves as more training data become available as exploration progresses. However, toward the later stages of exploration, more difficult-to-find nodes remain unexplored, causing a gradual decline in precision. It is important to note that the superiority of each method in terms of exploration efficiency remains consistent across both Figure~\ref{fig:prec} and Figure~\ref{fig:coverage}.

\end{document}